\documentstyle[aps,preprint]{revtex}
\newcommand{\be}{\begin{equation}}
\newcommand{\ee}{\end{equation}}
\newcommand{\Be}{\begin{eqnarray}}
\newcommand{\Ee}{\end{eqnarray}}
\newcommand{\f}{\frac}

\draft
\begin{document}

\title{The traversable wormhole with classical scalar fields}
\author{Sung-Won Kim\footnote{e-mail: sungwon@mm.ewha.ac.kr} }
\address{ Department of Science Education and\\ 
Center for High Energy Astrophysics and Isotope Studies(CAIS) \\
Ewha Womans University\\
Seoul 120-750, Korea}
\author{Sang Pyo Kim\footnote{e-mail: sangkim@knusun1.kunsan.ac.kr} }
\address{ Department of Physics\\
Kunsan National University\\
Kunsan 573-701, Korea}

\maketitle


\begin{abstract}
We study the Lorentzian static 
traversable wormholes coupled to quadratic scalar fields. 
We also obtain the solutions of the scalar fields and matters
in the wormhole background and find that  
 the minimal size of the wormhole should be quantized
under the appropriate boundary conditions for the positive 
non-minimal  massive scalar field.
\end{abstract}

\newpage

One of the most important  issues in making a practically usable Lorentzian wormhole
 is just the traversability\cite{MT88,MTY88}.
To make a Lorentzian wormhole traversable, one has usually  used an exotic 
matter which violates the well-known energy conditions\cite{MT88}.
For instance, a wormhole in the inflating cosmological model still requires the 
exotic matter to be traversable and to maintain its shape\cite{R93}.
It is known that the vacuum energy of the inflating wormhole does not 
change the sign of the exoticity function.
A traversable wormhole in the Friedmann-Robertson-Walker(FRW) 
cosmological model, however, does not necessarily require the
exotic matter at the very early times\cite{K96}.  

In this paper, we investigate the compatibility of a static wormhole with a
minimal and a positive non-minimal scalar 
field. 
We also obtain the solutions to the matters and scalar fields 
in wormhole background. Furthermore, we find that 
the minimum size of wormhole
should be quantized when appropriate boundary conditions are imposed
for the non-minimal  massive scalar field.
  
Firstly, we study the simplest case of a static Lorentzian wormhole with a 
minimal massless scalar field.
The additional matter Lagrangian due to the scalar field is
given by
\be
{\cal L} = \f{1}{2}\sqrt{-g}g^{\mu\nu}\varphi_{;\mu}\varphi_{;\nu} 
\label{eq:lag}
\ee
and the equation of motion for $\varphi$ by
\be
\Box\varphi=0. \label{eq:wave}
\ee
The stress-energy tensor for $\varphi$ is obtained from Eq. (\ref{eq:lag}) as
\be
T_{\mu\nu}^{(\varphi)} = \varphi_{;\mu}\varphi_{;\nu}
-\f{1}{2}g_{\mu\nu}g^{\rho\sigma}\varphi_{;\rho}\varphi_{;\sigma}.
\label{eq:energy}
\ee
Now the Einstein equation has an additional stress-energy tensor
(\ref{eq:energy})
\be
G_{\mu\nu} = R_{\mu\nu}-\f{1}{2}g_{\mu\nu}R = 8\pi T_{\mu\nu}
= 8\pi(T_{\mu\nu}^{\rm (w)}+T_{\mu\nu}^{(\varphi)}),
\label{eq:ein}
\ee
where $T^{\rm (w)}_{\mu\nu} $ is the stress-energy tensor of the
background matter that makes the traversable wormhole.
Assuming a spherically symmetric spacetime, one finds
the components of $T_{\hat{\mu}\hat{\nu}}^{\rm (w)} $ in orthonormal 
coordinates
\be
T^{\rm (w)}_{\hat{t}\hat{t}} = \rho(r,t),
~~T^{\rm (w)}_{\hat{r}\hat{r}} = -\tau(r,t),
~~T^{\rm (w)}_{\hat{\theta}\hat{\theta}} = P(r,t),
\label{eq:matter}
\ee
where $\rho(r,t), \tau(r,t)$ and $P(r,t)$ are  the 
mass energy density, radial tension per unit area, and lateral 
pressure, respectively, as measured by an observer at
a fixed $r, \theta, \phi$.

The metric of the static wormhole is given by
\be 
ds^2 = -e^{2\Lambda(r)}dt^2 +
\f{dr^2}{1-b(r)/r}
+ r^2 (d\theta^2+\sin^2\theta d\phi^2). 
\label{eq:metric}
\ee
The arbitrary functions $\Lambda(r)$ and $b(r)$ are lapse and 
wormhole shape functions, respectively.  The shape of the wormhole
is determined by $b(r)$. Beside the
spherically symmetric and static spacetime,  
we further assume a zero-tidal-force as seen by
stationary observer, $\Lambda(r)=0$,  to make the problem simpler. 
Thus not only the scalar field $\varphi$ but also the matter 
$\rho, \tau,$ and $P$ are assumed to depend only on $r$.
The components of $T_{\mu\nu}^{(\varphi)} $ in the static wormhole metric 
(\ref{eq:metric})  have the form
\Be
T_{tt}^{(\varphi)}&=&\f{1}{2}\left(1-\f{b}{r}\right)\varphi'^2 
\label{eq:scem1}\\
T_{rr}^{(\varphi)}&=&\f{1}{2}\varphi'^2 \label{eq:scem2}\\
T_{\theta\theta}^{(\varphi)}&=&-\f{1}{2}r^2\left(1-\f{b}{r}\right)
\varphi'^2 \label{eq:scem3}\\
T_{\phi\phi}^{(\varphi)}&=&
-\f{1}{2}r^2\left(1-\f{b}{r}\right)\varphi'^2\sin^2\theta, 
\label{eq:scem4}
\Ee
where and hereafter a prime denoted the differentiation with
respect to $r$.  In the spacetime with  the metric (\ref{eq:metric})
and $\Lambda = 0$, the field equation of $\varphi$ becomes
\be
\f{\varphi''}{\varphi'}+\f{1}{2}\f{(1-b/r)'}{(1-b/r)}+\f{2}{r} = 0 \quad
\quad \mbox{or} \quad \quad
r^4\varphi'^2\left(1-\f{b}{r}\right) = \mbox{const} ,
\label{eq:phi}
\ee
and the Einstein equations are given explicitly by
\Be
\f{b'}{8\pi r^2}&=& \rho(r,t) + \f{1}{2}\varphi'^2\left(1-\f{b}{r}\right), 
\label{eq:scein1}\\
\f{b}{8\pi r^3}&=& \tau(r,t) - \f{1}{2}\varphi'^2\left(1-\f{b}{r}\right), 
\label{eq:scein2}\\
\f{b-b'r}{16\pi r^3}  &=& P(r,t) - \f{1}{2}\varphi'^2\left(1-\f{b}{r}\right).
\label{eq:scein3}
\Ee
By redefining the effective matters by
\Be
\rho_{\rm eff}&=&\rho + \f{1}{2}\varphi'^2\left(1-\f{b}{r}\right),
\label{eq:scmatt1}\\
\tau_{\rm eff}&=&\tau - \f{1}{2}\varphi'^2\left(1-\f{b}{r}\right),
\label{eq:scmatt2}\\
P_{\rm eff}&=&P - \f{1}{2}\varphi'^2\left(1-\f{b}{r}\right),
\label{eq:scmatt3}
\Ee
we are able to rewrite the Einstein equations as
\Be
\f{b'}{8\pi r^2}  &=& \rho_{\rm eff}, \label{eq:sceff1}\\
\f{b}{8\pi r^3}  &=& \tau_{\rm eff}, \label{eq:sceff2}\\
\f{b-b'r}{16\pi r^3}  &=& P_{\rm eff}, \label{eq:sceff3}
\label{eq:sceff4}
\Ee
Thus one sees that  the conservation law of the 
effective 
stress-energy
tensor $T^{(\rm w)}_{\mu\nu}+
T^{(\varphi)}_{\mu\nu}$ still obeys the same equation
\be
\tau'_{\rm eff}+\f{2}{r}(\tau_{\rm eff}+P_{\rm eff}) = 0.
\ee

We now find the solutions of scalar field and
matter.
To determine the spatial distributions of $b(r), \rho(r), \tau(r), P(r)$,
and $\varphi(r)$, we need one more condition for them such as
the equation of state, $P_{\rm eff}=\beta\rho_{\rm eff}$.
With the appropriate asymptotic flatness imposed
we find the effective matter as functions of $r$ \cite{K96}
\Be
\rho_{\rm eff}(r) &\propto& r^{-2(1+3\beta)/(1+2\beta)}, \label{eq:sol1}\\
\tau_{\rm eff}(r) &\propto& r^{-2(1+3\beta)/(1+2\beta)}, \label{eq:sol2}\\
b(r) &\propto& r^{1/(1+2\beta)} , \qquad \beta < -\f{1}{2}.
\label{eq:sol3}
\Ee
Since the scalar field effects changes just $r^{-4}$ by the field equation
(\ref{eq:phi}), the matter is given by 
\Be
\rho &\sim& r^{-2(1+3\beta)/(1+2\beta)}+r^{-4}, \\
\tau &\sim& r^{-2(1+3\beta)/(1+2\beta)}+r^{-4}.
\Ee
Once $b(r)$ is known to depend on a specific value of $\beta$, 
we can integrate Eq. (\ref{eq:phi}) to obtain
\be
\varphi(r) \propto \int \f{dr}{r^2\sqrt{\left(1-b(r)/r\right)}}.
\label{eq:sol}
\ee
For example, when $b=b_0^2/r$, where  $\beta=-1$ and $b_0$ is the
minimum size of the throat of the wormhole, we obtain that
\be
\varphi = \varphi_0\left[1-\arccos\left(\f{b_0}{r}\right)\right] 
\ee
by assuming the boundary condition
that $\varphi > 0, ~\varphi|_{r=b_0} = \varphi_0,~$ and  
$\lim_{r\rightarrow\infty}\varphi=0$.  Thus the scalar field
decreases monotonically, i.e. $\varphi' < 0$,
and the matter has 
$\rho, \tau, P \propto r^{-4}$.

Secondly we consider a general quadratic scalar field,
whose stress-energy tensor is given by
\Be
T_{\mu\nu}&=&(1-2\xi)\varphi_{;\mu}\varphi_{;\nu}+
(2\xi-\f{1}{2})g_{\mu\nu}\varphi_{;\alpha}\varphi^{;\alpha}
-2\xi\varphi\varphi_{;\mu\nu}+2\xi g_{\mu\nu}\varphi\Box\varphi \nonumber\\
&&+\xi\left(R_{\mu\nu}-\f{1}{2}g_{\mu\nu}R\right)\varphi^2
-\f{1}{2}m^2 g_{\mu\nu}\varphi^2,
\Ee
where $\xi=0$ for minimal coupling and $\xi=\f{1}{6}$ for
conformal coupling.  The mass of the field is given by $m$.
The field equation for $\varphi$ is
\be
(\Box - m^2 - \xi R ) \varphi = 0.
\ee
For a non-minimal coupling there is a curvature
effect of wormhole background to  $T_{\mu\nu}$ and   $\varphi$. 
The scalar curvature in the metric (\ref{eq:metric}) with $\Lambda=0$
is given by
\be
R = \f{2b'}{r^2}
\ee
and 
the components of the stress-energy tensor for the scalar field by
\Be
T_{tt}^{(\varphi)} &=& (1-4\xi)\left[\f{1}{2}\left(1-\f{b}{r}\right)\varphi'^2
+ \left(\f{1}{2}m^2+\xi\f{b'}{r^2}\right)\varphi^2\right],\\
T_{rr}^{(\varphi)} &=& \f{1}{2}\varphi'^2 + \xi\f{4}{r} \varphi'\varphi
- \left(1-\f{b}{r}\right)^{-1} \left( \xi \f{b}{r^3}+\f{1}{2}m^2\right)
\varphi^2, \\
T_{\theta\theta}^{(\varphi)}&=&\left(1-\f{b}{r}\right)\left[\left(2\xi-
\f{1}{2}\right)r^2\varphi'^2 - 2\xi r \varphi'\varphi \right] \nonumber\\
&&+ \left[ 2\xi r^2\left( m^2 + \xi\f{2b'}{r^2}\right) + \xi
\left(\f{b}{2r}-\f{b'}{2}\right)-\f{1}{2}m^2r^2\right]\varphi^2, \\
T_{\phi\phi}^{(\varphi)}&=&T_{\theta\theta}^{(\varphi)}\sin^2\theta.
\Ee
As in the minimally-coupled case,
we are also able to find the redefinition of the effective 
matter
\Be
\rho_{\rm eff} &=& \rho +
(1-4\xi)\left[\f{1}{2}\left(1-\f{b}{r}\right)\varphi'^2
+ \left(\f{1}{2}m^2+\xi\f{b'}{r^2}\right)\varphi^2\right], \\
\tau_{\rm eff} &=& \tau - \f{1}{2}\varphi'^2\left(1-\f{b}{r}\right)
-\xi\f{4}{r}\left(1-\f{b}{r}\right)\varphi'\varphi
+\left(\xi\f{b}{r^3}+\f{1}{2}m^2\right)
\varphi^2, \\
P_{\rm eff} &=& P + \left(1-\f{b}{r}\right)\left[\left(2\xi-
\f{1}{2}\right)\varphi'^2 - \xi\f{2}{r} \varphi'\varphi \right] \nonumber\\
&&+ \left[ 2\xi \left( m^2 + \xi\f{2b'}{r^2}\right) + \xi
\left(\f{b}{2r^3}-\f{b'}{2r^2}\right)-\f{1}{2}m^2\right]\varphi^2.
\Ee

One has the same solution to the equation of the effective matter as  
Eqs.(\ref{eq:sol1}-\ref{eq:sol3}) with the same equation of state.
However, one has a more complicated scalar field equation
\be
\left(1-\f{b}{r}\right)\varphi''+\left(1-\f{b}{r}\right)'\varphi'
+\f{2}{r}\left(1-\f{b}{r}\right)\varphi'=
\left(m^2 + 2\xi\f{b'}{r^2}\right)\varphi
\label{eq:geneq}
\ee
>From the wormhole shape function
$b \sim r^{1/(1+2\beta)}$ with the same equation of state
$P_{\rm eff}=\beta\rho_{\rm eff}$, the asymptotic form of the field 
can be calculated near the throat and at the infinity. 
\Be
\varphi &\sim& \f{\exp{[km^2 r^{2(1+3\beta)/(1+2\beta)}]}}{r^{-2\xi/\beta}}
\sim \f{\exp{(km^2/\rho_{\rm eff})}}{r^{-2\xi/\beta}} \qquad {\rm at}~~ 
r \rightarrow b \nonumber \\
&\sim & e^{-mr}\qquad\qquad\qquad\qquad {\rm at}~~ r \rightarrow \infty,
\Ee
where $k=(1+2\beta)^2/[2\beta(1+3\beta)]b_0^{-2\beta/(1+2\beta)}$.  For the 
special case of $\beta=-1$, i.e., $b=b_0^2/r$, 
the scalar field has the asymptotic form near the throat
\be
\varphi \sim \f{\exp{(km^2r^4)}}{r^{2\xi}},\qquad\qquad {\rm at}~ 
r\rightarrow b
\ee
where $k \rightarrow 1/4b_0$.
The scalar field begins to increase from the throat
very rapidly with $r$, but
decreases exponentially at the infinity.

To find the exact solution to the scalar field, 
we rewrite the field equation (\ref{eq:geneq}) as
\be
\f{d^2\varphi}{ds^2}-r^4(m^2+\xi R)\varphi = 
\f{d^2\varphi}{ds^2}-f(s)\varphi=0,
\label{eq:Sch}
\ee
where $s=\int r^{-2}(1-b/r)^{-1/2}dr$ and 
$f(s)=r^2(s)(m^2+2\xi b'(s)/r^2(s))$.
One may interpret Eq.~(\ref{eq:Sch}) as a Schr\"odinger equation
with the zero energy, 
which is, however,  not easy to solve in general.  
When $m=\xi=0$, the field equation becomes just the 
minimal  massless case and the solution is $\varphi \sim s$,
which is given by Eq.~(\ref{eq:sol}).
For the general case,  we can make use of the analogy
with a bounded potential problem in the region
$ r_0 \le r < \infty $ or $ 0 \le s \le s_0 $, where $r_0$ is the place
of the throat.
In this region, we can find the asymptotic values of potential 
$f(s)$ and $\varphi$ as  
\be
\lim_{r \rightarrow r_0  ~{\rm or}~ s \rightarrow 0} f(s)  < 0, 
\qquad \lim_{r \rightarrow r_0  ~{\rm or}~ s \rightarrow 0} \varphi \sim 0,
\ee  
\be
\lim_{r \rightarrow \infty ~{\rm or}~ s \rightarrow s_0} f(s)  = \infty, 
\qquad \lim_{r \rightarrow \infty ~{\rm or}~ s \rightarrow s_0} \varphi 
\sim 0,
\ee
if $m^2 + 2\xi b'(s) / r^2 < 0 $ near throat or $ m^2 r^2_0 + 2\xi b'(0) < 0$.
Otherwise, there is no solution to $\varphi$.
The bounded potential shows that one parameter should be
quantized.  Which parameter is quantized?

We consider the specific case of $b=b^2_0/r$ as above.
Then the field equation  becomes
\be
\f{d^2\varphi}{ds^2}+[2\xi b^2_0 - m^2 b^2_0 \sec^4(b_0s)]\varphi=0,
\label{sch}
\ee
where $s=(1/b_0)\arccos(b_0/r)$.  
First, the massless scalar field has the solution
\be
\varphi=\varphi_0\cos(\sqrt{2\xi}b_0s)
=\varphi_0\cos\left[\sqrt{2\xi}\arccos\left(\f{b_0}{r}\right)\right].
\ee
Second, in the massive case, however, 
one has  the energy parameter $E=2\xi b_0^2$ and the potential 
$V=m^2b^2_0\sec^4(b_0s)$. The relation $E>V$ or
$2\xi>m^2b^2_0$ must be satisfied to guarantee a bounded solution. 
For small $s$, that is the region near throat, Eq.~({\ref{sch}) is 
approximately the harmonic oscillator problem
\be
\f{d^2\varphi}{dx^2}+(\lambda-x^2)\varphi=0
\ee
with $x^2=2mb^3_0s^2$ and $\lambda=(2\xi-m^2b^2_0)/(2mb_0)$.  The solution is
the harmonic wave functions
$\varphi_n=e^{-x^2/2}H_n(x)$, the Hermite polynomial. 
>From the energy quantization, we get
$mb_0=2n+1+\sqrt{(2n+1)^2+2\xi}$, where $n$ is odd number. 
>From this result we may conclude 
that $\xi b^2_0$ or the minimal size of the wormhole
should be quantized for the non-minimal 
positive coupling $\xi > 0$, including the conformal
coupling $\xi = 1/6$.

In this paper we found the solutions of the wormhole with
minimal and non-minimal scalar fields. 
For a positive non-minimal massive scalar field case
we find that the size of throat,
the minimal size of the wormhole, should be quantized
in order to have the scalar field solution satisfying the appropriate
boundary conditions.
We also find the solutions to the matter and the scalar 
field in each cases.

\acknowledgements

This work was supported in part by Non-directed program of Ministry 
of Education, 1994, in part by the Basic Science Research 
Institute in Ewha Womans University, BSRI-97-2427, and in part by
the Korea Science and Engineering Foundation No. 95-0702-04-01-3.

\end{document}